\documentclass{article}

\usepackage{amsmath}

\begin{document}

\title{Comment on ``Singularity-free Cosmological Solutions with 
Non-rotating Perfect Fluids''}

\author{L. Fern\'andez-Jambrina\footnote{E.T.S.I. Navales, Universidad Polit\'ecnica de Madrid,  
Arco de la Victoria s/n, E-28040 Madrid, Spain; e-mail: 
lfernandez@etsin.upm.es; URL: http://debin.etsin.upm.es/ilfj.htm}}

\date{\today}
\maketitle

\begin{abstract}
A conjecture stated by Raychaudhuri which claims that 
the only physical perfect fluid non-rotating non-singular cosmological models are comprised in the 
Ruiz-Senovilla and Fern\'andez-Jambrina families is shown to be incorrect. An explicit 
counterexample is provided and the failure of the argument leading to 
the result is explicitly pointed out. \end{abstract}

Since the publication of the first non-singular cosmological model 
with a realistic equation of state \cite{seno}, much effort has been 
devoted to either produce new regular models or to prove that they 
are a set of measure zero in some sense. Raychaudhuri \cite {ray} attempts 
to settle the issue by proving the following conjecture:\\

\noindent \textbf{Conjecture:} The only solutions to Einstein equations 
that fulfill the following conditions,

\begin{enumerate}
    \item  Non-singularity: The curvature and physical scalars are 
    regular in the whole spacetime and do not blow up at infinity.

    \item  Non-rotation: The vorticity of the cosmological fluid is 
    zero.
    
    \item  Perfect fluid: The matter content of the spacetime is a 
    perfect fluid. Therefore the energy-momentum is 
    $T=(p+\rho)u\otimes u+p\,g$, where $u$ is the velocity of the 
    fluid, $p$ is the pressure, $\rho$ is the density and $g$ is the 
    metric.

    \item  Cosmology: There is fluid throughout the space which 
    fulfills the energy conditions $0\le p\le \rho$. Discontinuities are excluded.

    \item  $\frac{\partial p}{\partial\rho}$ is positive everywhere.
    
    \item The acceleration of the fluid is hypersurface-orthogonal.
\end{enumerate} 
are those included in the Ruiz-Senovilla  \cite{ruiz} and 
Fern\'andez-Jambrina \cite{leo} families. 

An infinite family of counter-examples for this claim is supported by the 
models in \cite{wide} and \cite{stiff},
\begin{equation}
ds^2=e^{2K}(-dt^2+dr^2)+e^{-2U}dz^2+\rho^2e^{2U}d\phi^2,\label{metric}
\end{equation}
which correspond to a cylindrical cosmological model with matter 
content due to a stiff perfect fluid, $\rho=p$,
\begin{equation}
    p=\alpha e^{-2K},
\end{equation} with $\alpha=\mathrm{const}.>0$ 

The coordinates are comoving since the velocity of the fluid is just
\begin{equation} {u}= {\rm e}^{ -\frac{1}{2}\,K}\partial_t.
 \end{equation}

The metric function $U$ is a solution of the reduced two-dimensional wave 
equation,
 \begin{subequations}
 \label{einstein} 
 \begin{eqnarray}
 U_{tt}-U_{rr}-\frac{U_{r}}{r}=0,\label{U2}
 \\
 K_{t}=U_{t}+2r U_{t}U_{r},\label{Kr2}
 \\
 K_{r}=U_{r}+r(U_{t}^2+U_{r}^2)+\alpha r,\label{Kt2}
 \end{eqnarray}
 \end{subequations}
and $K$ is obtained by a quadrature when $U$ is known.

Therefore, the general solution for this problem may be obtained from
the Cauchy problem for the wave equation, for initial data
$U(r,0)=f(r)$, $U_{t}(r,0)=g(r)$, 

\begin{equation}\label{cauchy}
U(r,t)=\frac{1}{2\pi}\int_{0}^{2\pi}\!\!d\phi\int_0^1\!d\tau 
 \frac{\tau}{\sqrt{1-\tau^2}}\left\{tg(v)+f(v)+tf'(v) 
 \frac{t\tau^2+r\tau\cos\phi }{v}
\right\},\end{equation}
where $v=\sqrt{r^2+t^2\tau^2+2rt\tau\cos\phi}$.

These models are non-singular provided
$U|_{r=0}$ does not decrease too fast for 
large values of $|t|$,

\begin{equation}
U(0,t)=\int_0^1d\tau 
 \frac{\tau}{\sqrt{1-\tau^2}}\left\{tg(|t|\tau)+f(|t|\tau)+|t|\tau 
 f'(|t|\tau) 
\right\}\ge -\frac{1}{2}\ln |t|+b.\label{req}\end{equation} 
 
The fluid invariants can be shown to be regular and vanish at spatial and time 
infinity. The same happens with the curvature invariants.

It is obvious that these simple models fulfill Raychaudhuri's 
requirements: They are non-singular and non-rotating, the pressure and 
the density are positive at every point of the spacetime and they are 
related by a state of equation. 

However, they do not belong to the Ruiz-Senovilla or
Fer\-n\'an\-dez-Jam\-bri\-na family.  In fact, they are a generalization of the
latter family.

What is wrong then in Raychaudhuri's result? The most obvious 
failure in the reasoning leading to his claim lies at the onset of his line 
of thought. 

The author claims that the line of maxima described by 
$\frac{\partial{p}}{\partial{r}}=0$ is a constant $r$ line and 
similarly the line of maxima described by 
$\frac{\partial{p}}{\partial{t}}=0$ is a constant $t$ line. The main 
assumption for such results is that the integrals of 
$\frac{\partial^2{p}}{\partial{r^2}}$, $\frac{\partial^2{p}}{\partial{t^2}}$ along an infinite path are 
necessarily infinite, but this is not true if the pressure decreases 
fast enough close to infinity. 

This happens, for instance, when the pressure decreases 
exponentially. A simple and integrable example is provided by the 
solution of the Cauchy problem to the reduced wave equation with 
initial data $f(r)=\beta x^4$, $g(x)=\gamma x^2$, $\beta>0$,

\begin{subequations}
\begin{eqnarray}
U(r,t)=\frac{2}{3}\gamma t^3+\gamma tr^2+\frac{8}{3}\beta t^4+8\beta t^2r^2+
\beta r^4,\\
K(r,t)={\frac {512}{9}} {\beta}^{2}{r}^{2}{t}^{6}+{\frac {64}{3}} \beta\gamma  {r}^{2}
{t}^{5}+ \left( {\frac {448}{3}} {\beta}^{2}{r}^{4}+\frac{8}{
3} \beta+2 {\gamma}^{2}{r}^{2} \right) {t}^{4}\nonumber\\+ \left( \frac{2}{3} \gamma+{
\frac {112}{3}}\beta \gamma {r}^{4} \right) {t}^{3}+ \left( 8 \beta
 {r}^{2}+2 {\gamma}^{2}{r}^{4}+64 {\beta}^{2}{r}^{6} \right) {t}^{2
}\nonumber\\+ \left( 8 \beta\gamma  {r}^{6}+\gamma {r}^{2} \right) t+\frac{1}{6} {
\gamma}^{2}{r}^{6}+\beta {r}^{4}+2 {\beta}^{2}{r}^{8}+\frac{1}{2}\alpha
 {r}^{2},
\end{eqnarray}
\end{subequations}
and the subsequent quadrature for $K$.

If we analyze the lines of maxima for $p=\alpha e^{-2K}$ we find
\begin{eqnarray}
    \frac{1}{p}\frac{\partial p}{\partial r}=-\frac{2}{9}r
\left(1024 {\beta}^{2}{t}^{6}+384 \beta  \gamma{t}^{5}+ \left( 5376 {
\beta}^{2}{r}^{2}+36 {\gamma}^{2} \right) {t}^{4}\right.\nonumber\\ \left.+1344\beta\gamma {r}^{2}
  {t}^{3} + \left( 144 \beta+72 {\gamma}^{2}{r}^{2}+3456 {
\beta}^{2}{r}^{4} \right) {t}^{2}+ \left( 18 \gamma+432 \beta\gamma{r}^{4}
  \right) t\right.\nonumber\\ \left.+9 \alpha+9 {\gamma}^{2}{r}^{4}+36 \beta {r}^{2}
+144 {\beta}^{2}{r}^{6}\right),
\end{eqnarray}
that indeed the $r=0$ line provides a maxima. On the other hand,

\begin{eqnarray}
    \frac{1}{p}\frac{\partial p}{\partial t}=-\frac{2}{3}
    \left( 8 \beta {r}^{4}+32 \beta {t}^{2}{r}^{2}+4 \gamma t{r}^{2}+1 \right) 
    \nonumber\\ \left( 6 \gamma {t}^{2}+3 \gamma {r}^{2}+32 \beta {
t}^{3}+48 \beta t{r}^{2} \right),
\end{eqnarray} 
it is clear that there is no constant $t$ line of maxima.

This solution does not therefore exhibit any simple property of separability. 
Furthermore, it also contradicts the claim in section 9 about  space 
time reversibility of non-singular solutions. It is explicit that this 
solution is not time reversible. In fact, the models in \cite{wide} 
are generically non-reversible.

Finally, it is to be pointed out that the example in this work is 
just an easy integrable case. The features exhibited by this model are 
shared by the whole family, except for the most simple solutions, 
like \cite{leo}.

\section*{Acknowledgements}
The present work has been supported by Direcci\'on General de
Ense\~nanza Superior Project PB98-0772. L.F.J. wishes to thank
 F.J. Chinea, L.M. Gonz\'alez-Romero, F. Navarro-L\'erida and  M.J. Pareja 
for valuable discussions.


\begin{thebibliography}{99}
\bibitem{seno} J.M.M. Senovilla, \textit{ Phys. Rev. Lett.} \textbf{
64}, 2219 (1990).

\bibitem{ray} A.K. Raychaudhuri, \textit{Gen. Rel. Grav.} \textbf{36}, 
343-359 (2004).

\bibitem{ruiz} E. Ruiz, J.M.M. Senovilla, \textit{ Phys. Rev.} \textbf{ 
D45}, 1995 (1992). 

\bibitem{leo}L. Fern\'andez-Jambrina, \textit{ Class. Quantum Grav.} \textbf{14}, 3407 (1997).

\bibitem{wide} L. Fern\'andez-Jambrina, L.M. Gonz\'alez-Romero, \textit{Phys. Rev.} \textbf{
D 66}, 024027 (2002).    

\bibitem{stiff} L. Fern\'andez-Jambrina, L.M. Gonz\'alez-Romero, 
\textit{Journ. Math. Phys.} \textbf{45}, 2113 (2004).


\bibitem{manolo}L. Fern\'andez-Jambrina, L.M. Gonz\'alez-Romero, 
\textit{Class. Quantum Grav.} \textbf{16}, 953 (1999).
\end{thebibliography}
\end{document}